\documentclass[aps,a4paper,superscriptaddress,nofootinbib,twocolumn]{revtex4-1}
\usepackage{amsmath,amssymb,gensymb,graphicx,multirow,dcolumn,bm,latexsym,soul,nicefrac,acronym}
\usepackage[mathscr]{euscript}
\usepackage{appendix}
\usepackage[colorlinks,linkcolor=blue,citecolor=blue,urlcolor=blue ]{hyperref}
\usepackage{float}
\usepackage{soul}
\usepackage{verbatim}
\usepackage{color}
\usepackage{acronym}
\usepackage{subfigure}
\usepackage{appendix}
\usepackage{longtable}
\usepackage{footnote}
\usepackage{cancel}
\usepackage{ulem} 

\newcommand{\mnras}{Monthly Notices of the Royal Astronomical Society}

\newcommand{\be}{\begin{equation}}
\newcommand{\ee}{\end{equation}}
\newcommand{\bea}{\begin{eqnarray}}
\newcommand{\eea}{\end{eqnarray}}
\newcommand{\bw}{\begin{widetext}}
\newcommand{\ew}{\end{widetext}}
\newcommand{\nn}{\nonumber}

\newcommand{\eq}[1]{Eq.~(\ref{#1})}
\newcommand{\fig}[1]{Fig.~\ref{#1}}
\newcommand{\tab}[1]{Table.~\ref{#1}}

\newcommand{\TRC}{MOE Key Laboratory of TianQin Mission, TianQin Research Center for Gravitational Physics $\&$  School of Physics and Astronomy, Frontiers Science Center for TianQin, CNSA Research Center for Gravitational Waves, Sun Yat-sen University (Zhuhai Campus), Zhuhai 519082, China}

\begin{document}
\title{Revisiting Stochastic Gravitational Wave Background in the Strong Signal Case}
\author{Zheng-Cheng Liang}
\affiliation{\TRC}
\author{Zhi-Yuan Li}
\affiliation{\TRC}
\author{En-Kun Li}
\affiliation{\TRC}
\author{Jian-dong Zhang}
\affiliation{\TRC}
\author{Yi-Ming Hu}
\email{huyiming@sysu.edu.cn}
\affiliation{\TRC}

\date{\today}

\begin{abstract}
Weak-signal limit is often used in estimating stochastic gravitational-wave background (SGWB) intensities. 
This approximation fails and the signal-to-noise ratio (SNR) can be much weaker when background signals are loud compared to the detector noise. 
In this work, we find that this issue is especially significant when dealing with networks of detectors that are widely separated. 
For the TianQin + LISA network, the SNR estimated under the weak-signal limit might be off by as large as an order of magnitude. 
Contour plots of SNR over the parameter spaces are also presented to indicate regions that will be affected by this correction. 
Our results suggest that DA and DB type extragalactic double white dwarfs may yield an SGWB with SNR surpassing 100 after 1 year of operation in weak-signal limit scenario, with a redshift-independent merger rate of about $500\,\,{\rm Mpc^{-3}\,Myr^{-1}}$. 
In fact, this value falls significantly below the necessary threshold. 
Similar influences arise for first-order phase transitions, yet pinning down the detectable parameters remains formidable due to model uncertainties. 
\end{abstract}

\keywords{}

\pacs{04.25.dg, 04.40.Nr, 04.70.-s, 04.70.Bw}

\maketitle
\acrodef{SGWB}{stochastic \ac{GW} background}
\acrodef{GW}{gravitational-wave}
\acrodef{CBC}{compact binary coalescence}
\acrodef{MBHB}{supermassive black hole binary}
\acrodef{BBH}{binary black hole}
\acrodef{EMRI}{extreme-mass-ratio inspiral}
\acrodef{DWD}{double white dwarf}
\acrodef{GDWD}{Galactic double white dwarf}
\acrodef{EDWD}{extragalactic double white dwarf}
\acrodef{BH}{black hole}
\acrodef{NS}{neutron star}
\acrodef{BNS}{binary neutron star}
\acrodef{LIGO}{Laser Interferometer Gravitational-Wave Observatory}
\acrodef{LISA}{Laser Interferometer Space Antenna}
\acrodef{TQ}{TianQin}
\acrodef{KAGRA}{Kamioka Gravitational Wave Detector}
\acrodef{ET}{Einstein telescope}
\acrodef{DECIGO}{DECi-hertz interferometry GravitationalWave Observatory}
\acrodef{CE}{Cosmic Explorer}
\acrodef{NANOGrav}{The North American Nanohertz Observatory for Gravitational Waves}
\acrodef{LHS}{left-hand side}
\acrodef{RHS}{right-hand side}
\acrodef{ORF}{overlap reduction function}
\acrodef{ASD}{amplitude spectral density}
\acrodef{PSD}{power spectral density}
\acrodef{SNR}{signal-to-noise ratio}
\acrodef{TDI}{time delay interferometry}
\acrodef{PIS}{peak-integrated sensitivity}
\acrodef{PLIS}{power-law integrated sensitivity}
\acrodef{GR}{general relativity}
\acrodef{PBH}{primordial black hole}
\acrodef{SSB}{solar system baryo}
\acrodef{PT}{phase transition}
\acrodef{SM}{Standard Model}
\acrodef{EWPT}{electroweak phase transition}
\acrodef{CPTA}{Chinese Pulsar Timing Arrays}
\acrodef{PTA}{Pulsar Timing Arrays}
\acrodef{RD}{radiation-dominated}
\acrodef{MD}{matter-dominated}
\acrodef{NG}{Nambu-Goto}

\section{Introduction}
The \ac{SGWB} is a superposition of independent and unresolvable \ac{GW} from diverse sources, with the astrophysical component including \ac{BBH}, \ac{BNS}, \ac{DWD}, etc~\cite{Christensen:2018iqi,Romano:2019yrj}. 
Alternatively, the cosmological component traces back to early Universe processes, such as inflation, first-order \acp{PT}, and cosmic defects~\cite{deAraujo:2000gw,Martinovic:2020hru}. 
This diversity enables the \ac{SGWB} to cover multiple frequency ranges. 
A recent significant milestone was made by collaborations utilizing \ac{PTA}, offering compelling evidence for the existence of \ac{SGWB} in the nanoHertz (nHz) band~\cite{NANOGrav:2023gor,Xu:2023wog,EPTA:2023fyk,Reardon:2023gzh}. 
While ground-based \ac{GW} detectors have yet to detect the \ac{SGWB} in the hundred-hertz frequency band, the upper limit of \ac{SGWB} intensity in this range has been estimated based on available observation data~\cite{KAGRA:2021kbb}. 
In the future, with the potential advancements in space-borne \ac{GW} detectors, \ac{SGWB} could be detected within the millihertz (mHz) frequency band~\cite{TianQin:2015yph,LISA:2017pwj}.

Space-borne \ac{GW} detectors, like TianQin~\cite{TianQin:2015yph}, \ac{LISA}~\cite{LISA:2017pwj}, are typically composed of three satellites arranged in an equilateral triangle formation. 
However, the unequal armlengths resulting from detector motion present a challenge in canceling laser noise, leading to the requirement of \ac{TDI} technique~\cite{Tinto:1999yr,Tinto:2020fcc}. 
Among the various \ac{TDI} combinations, the unequal-arm Michelson with the noise-orthogonal basis (A,E,T) is commonly employed. 
The A/E channels are specifically tailored for detecting \ac{GW}, providing optimized detection sensitivity~\cite{Vallisneri:2007xa}. 
The T channel exhibits a substantially lower sensitivity to \ac{GW} compared to the A/E channels, making it more suitable for monitoring detector noise rather than efficiently detecting \ac{GW}~\cite{Hogan:2001jn}.

One often relies on the auto- or cross-correlation of the channel data to derive the \ac{PSD} of \ac{SGWB}. 
The null-channel method utilizes the auto-correlation of the null channel to effectively track and isolate noises, allowing for the identification of the \ac{SGWB} component within a single space-borne detector~\cite{Tinto:2001ii,Hogan:2001jn,Adams:2013qma,Smith:2019wny,Boileau:2020rpg,Muratore:2021uqj,Cheng:2022vct}. 
While in scenarios where a network of space-borne detectors is deployed, each responding to a common \ac{SGWB} but with uncorrelated noises, distinguishing between the \ac{SGWB} and detector noise can be achieved through cross-correlating the channel data from two or more detectors~\cite{Hellings:1983fr,1987MNRAS.227..933M,Christensen:1992wi,Flanagan:1993ix,Allen:1997ad,Ungarelli:2001xu,LIGOScientific:2017zlf,Hu:2024toa}.

Following the null-channel and cross-correlation methods, the expectation of detection measurement is primarily determined by the \ac{PSD} of \ac{SGWB} and the \ac{ORF}, which quantifies the reduction in the correlation to an \ac{SGWB} resulting from the relative separation and orientation of the channels. 
On the other hand, the variance of detection measurement is solely influenced by the noise \ac{PSD}, given the \ac{SGWB} is significantly lower than the detector noise (weak-signal limit). 
Based on the expectation and variance, the detection \ac{SNR} can be calculated, taking into account factors such as correlation time accumulation and frequency band integration. 
Building upon the \ac{SNR} for the weak-signal limit, the \ac{PLIS} and \ac{PIS} sensitivity curve is derived and commonly used to showcase the detection capability for the \ac{SGWB}~\cite{Thrane:2013oya,Schmitz:2020syl}. 
Additionally, inversely solving the \ac{SNR} can be employed to constrain the detectable regions of source parameters, as exemplified by \ac{SNR} contour plots~\cite{Caprini:2015zlo,Kuroyanagi:2018csn,Caprini:2019egz}.

While the weak-signal limit is applicable in most scenarios, there may exist \ac{SGWB} with intensities matching or exceeding the detector noise
~\cite{Amaro-Seoane:2012vvq,Pan:2019uyn,Huang:2020rjf,Sharma:2020btq,Zhou:2022nmt,Wu:2023bwd,Torres-Orjuela:2023hfd,Mentasti:2023uyi,Song:2024pnk}. 
Therefore, Allen et al. extended the \ac{SNR} to handle arbitrarily large \ac{SGWB}, specifically tailored for an \ac{ORF}-invariant channel pair ~\cite{Allen:1997ad,Cornish:2001bb,Kudoh:2005as}. 
They highlighted that the existence of \acp{SGWB} can impact the variance of detection measurements, potentially resulting in an overestimation of detection capability~\cite{Allen:1997ad,Cornish:2001bb,Kudoh:2005as}. 
This \ac{SNR} estimator has been applied in a series of subsequent studies~\cite{Wang:2021njt,Brzeminski:2022haa,Liang:2022ufy,Hu:2024toa}. 
However, to the best of our knowledge, it has not been utilized for constructing \ac{PLIS} and \ac{PIS} sensitivity curves, nor in generating \ac{SNR} contour plots.

In this paper, we further develop a more robust \ac{SNR} estimation, adaptable to multiple channel pairs and time-varying \ac{ORF}. 
By incorporating the correction of the \ac{SNR} estimation, we revise the \ac{PLIS} and \ac{PIS} sensitivity curves. 
Focusing on TianQin, the TianQin I+II network, and the TianQin + LISA network, we identify gaps in sensitivity curves between the scenarios of approximation and non-approximation. 
Additionally, to illustrate the gaps in constraining detectable parameter regions between these two scenarios, we employ the \ac{SNR} contour plot.

The structure of the paper is as follows. 
In Sec.~\ref{sec:mission}, we introduce the space mission discussed in this study. 
Sec.~\ref{sec:formalism} and~\ref{sec:beyond} are dedicated to \ac{SGWB} detection with and without the weak-signal limit. 
The case study is presented in Sec.~\ref{sec:Case}. 
Finally, a brief conclusion and discussion is provided in Sec.~\ref{sec:conclusion}.

\section{Space mission}\label{sec:mission}
Several space missions have been proposed for \ac{GW} detection, such as TianQin~\cite{TianQin:2015yph}, LISA~\cite{LISA:2017pwj}, among others. 
This paper specifically centers on the TianQin and LISA missions.

\subsection{detector design}
\begin{figure}[t]
	\centering
	\includegraphics[height=5.3cm]{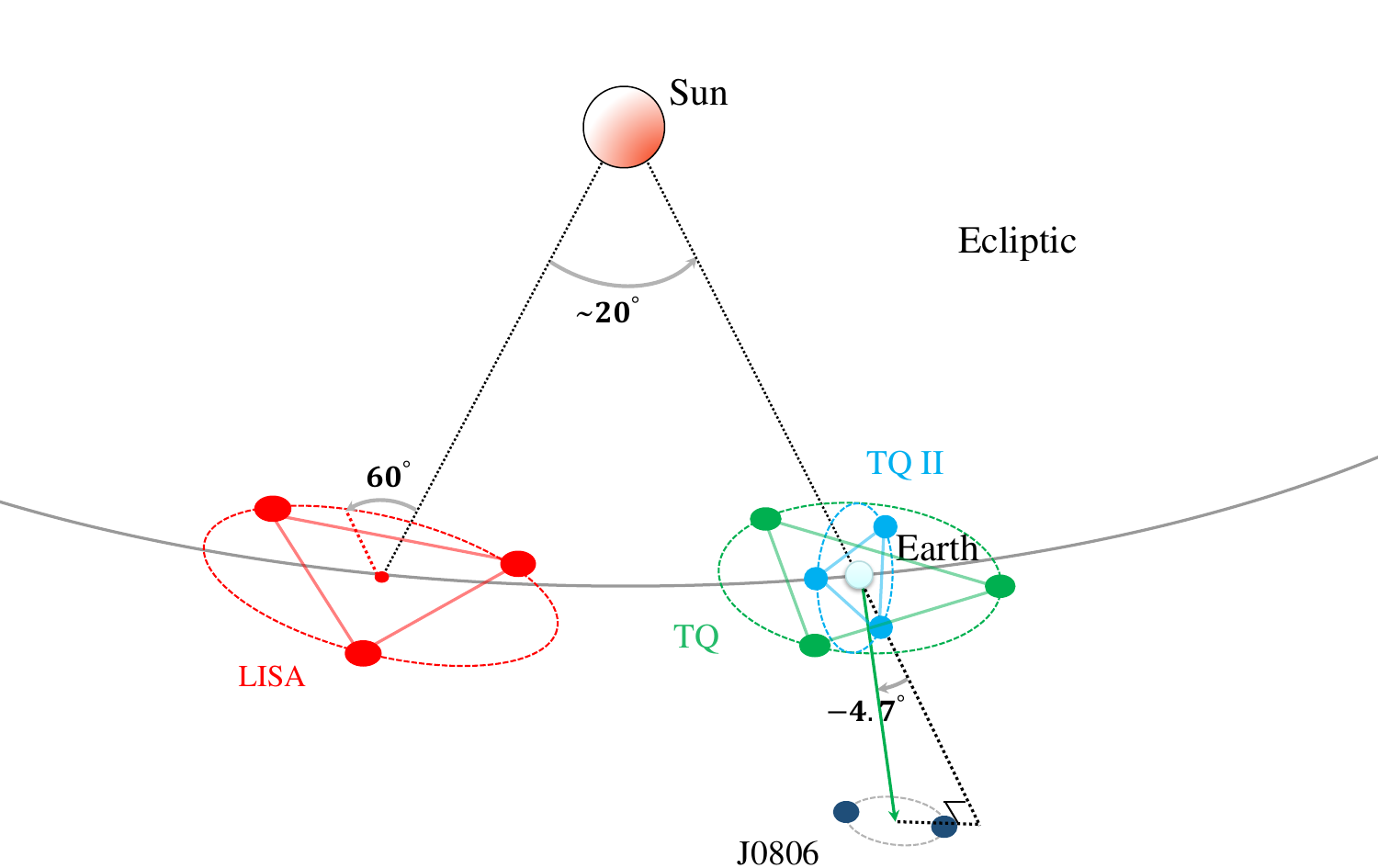}
	\caption{Schematic diagram of TianQin, TianQin II, and LISA in the ecliptic plane.}
	\label{fig:TQ_LISA}
\end{figure}

As shown in \fig{fig:TQ_LISA}, TianQin is a space-borne \ac{GW} detector comprising three identical satellites in Earth's orbit. 
These satellites will have a 3.64-day orbital period and will be positioned at a radius of approximately $10^{5}\,\, {\rm km}$.  When operational, they will form an equilateral triangle with an arm length ($L_{\rm TQ}$) of about $1.7\times10^{5}\,\, {\rm km}$. TianQin will operate on a ``three months on + three months off" schedule, with a proposed second mission, TianQin II, aimed at filling detection gaps~\cite{Ye:2019txh}. 
The orbital planes of TianQin and TianQin II are arranged perpendicular to each other, creating the TianQin I+II network.
The nominal work scheme leads to no overlap in the operation times of TianQin and TianQin II; we instead consider a scenario where the observation time can be extended and correspond to an enhanced operational mode of ``four months on + two months off".

LISA, on the other hand, is designed to orbit the Sun, trailing Earth by approximately $20\degree$ and maintaining a fixed $60\degree$ angle with the ecliptic plane. 
It will comprise three satellites separated by a distance of about $2\times10^{6}\,\,{\rm km}$, forming the armlength ($L_{\rm LISA}$) of the detector. 
LISA will share the similar operating period and detection band as TianQin, which opens up the opportunity to establish the TianQin + LISA network, allowing for a half-a-year overlap within 1 year. 

For convenience, shorthand notations can be used in figures and equations: TQ for TianQin, TQ II for TianQin II, TT for the TianQin I+II network, and TL for the TianQin + LISA network. 

\subsection{TDI channel}
Due to the detector motion, maintaining an equilateral shape for a space-borne detector is challenging, making it difficult to implement an equal-arm Michelson to cancel laser noise. 
In response to this challenge, the \ac{TDI} combination has been proposed~\cite{Tinto:1999yr}. 
The first-generation \ac{TDI} is initially designed for static detectors and includes components such as the Sagnac, unequal-arm Michelson, Relay, Monitor, and Beacon~\cite{Tinto:2020fcc}. 
This paper focuses on the unequal-arm Michelson (X,Y,Z), which can be established using each satellite and its adjacent link. Furthermore, the corresponding noise-orthogonal bases can be constructed by~\cite{Vallisneri:2007xa}
\bea
\label{eq:channel_AET}
\nn
{\rm A}&=&\frac{1}{\sqrt{2}}({\rm Z}-{\rm X}),\\
\nn
{\rm E}&=&\frac{1}{\sqrt{6}}({\rm X}-2{\rm Y}+{\rm Z}),\\
{\rm T}&=&\frac{1}{\sqrt{3}}({\rm X}+{\rm Y}+{\rm Z}).
\eea

Following the successful cancel of laser noise, secondary noise, including optical-path noise and acceleration noise, become more pronounced. 
Assuming stationary noise conditions, the \ac{PSD} for the \ac{TDI} channel (A, E, T) can be calculated by
\bw
\bea
\label{eq:Pn_AET}
\nn
P_{\rm n_{\rm A/E}}(f)
&=&\frac{2\sin^{2}\big[\frac{f}{f_{\ast}}\big]}{L^{2}}
\bigg[\bigg(\cos\big[\frac{f}{f_{\ast}}\big]+2\bigg)S_{\rm p}(f)+2\bigg(\cos\big[\frac{2f}{f_{\ast}}\big]+2\cos\big[\frac{f}{f_{\ast}}\big]
+3\bigg)\frac{S_{\rm a}(f)}{(2\pi f)^{4}}\bigg],\\
P_{\rm n_{T}}(f)
&=&\frac{8\sin^{2}\big[\frac{f}{f_{\ast}}\big]\sin^{2}\big[\frac{f}{2f_{\ast}}\big]}{L^{2}}
\bigg(S_{\rm p}(f)+4\sin^{2}\big[\frac{f}{2f_{\ast}}\big]\frac{S_{\rm a}(f)}{(2\pi f)^{4}}\bigg),
\eea
\ew
where the characteristic frequency $f_{*}=c/(2\pi L)$, $S_{\rm p}$ and $S_{\rm a}$ represent the \acp{PSD} of optical-path noise and acceleration noise, respectively. 
For more comprehensive information on the parameters of TianQin and LISA, readers are encouraged to refer to Refs.~\cite{TianQin:2020hid,Babak:2021mhe}. 
Unless otherwise stated, our study primarily focuses on the case with using the A/E/T channels. 

\section{Formalism}\label{sec:formalism}
\subsection{Stochastic background}
In the transverse-traceless gauge, the metric perturbation of an \ac{SGWB} can be expressed as a superposition of plane wave:
\bea
\label{eq:h_ab}
\nn
h(t,\vec{x})&=&\sum_{P=+,\times}\int_{-\infty}^{\infty}{\rm d}f\int_{S^{2}}{\rm d}\hat{\Omega}_{\hat{k}}\,
\widetilde{h}_{P}(f,\hat{k})\textbf{e}^{P}(\hat{k})\\
&&\times e^{{\rm i}2\pi f[t-\hat{k}\cdot\vec{x}(t)/c]},
\eea
where the wave vector of the gravitational wave is denoted by $\hat{k}$, the polarization tensor $\textbf{e}^{P}(\hat{k})$ refers to the polarization $P$ of the \ac{GW}, the speed of light is represented by $c$. 

To begin, we narrow our focus to the Gaussian-stationary, unpolarized, and isotropic \ac{SGWB}. 
Therefore, it is reasonable to assume that the \ac{SGWB} possesses a zero-mean Fourier amplitude $\widetilde{h}_{P}(f,\hat{k})$, with the direction-dependent \ac{PSD} $\mathscr{P}_{\rm h}$:
\be
\label{eq:Ph}
\langle\widetilde{h}_{P}(f,\hat{k})\widetilde{h}^{*}_{P'}(f',\hat{k}')\rangle
=\frac{1}{4}\delta(f-f')\delta_{PP'}\delta^{2}(\hat{k}-\hat{k}')\mathscr{P}_{\rm h}(|f|,\hat{k}).
\ee
Here the factor $1/4$ stems from both the definition of one-side \ac{PSD} and the average of polarization, $\delta_{ij}$ denotes the Kronecker delta. 
The \ac{PSD} of \ac{SGWB} incorporates contributions from all directions in the sky, as reflected in the all-sky integral of the $\mathscr{P}_{\rm h}$:
\bea
\label{eq:S2P}
\nn
S_{\rm h}(f)
&=&
\int_{S^{2}}{\rm d}\hat{\Omega}_{\hat{k}}
\mathscr{P}_{\rm h}(f,\hat{k})\\
&=&4\pi \mathscr{P}_{\rm h}(f)
\eea
where the second line of this equation is derived under the assumption of isotropy in the \ac{SGWB}. 

To further characterize the distribution of energy across different frequencies in the \ac{SGWB}, a commonly used quantity is the dimensionless energy spectrum density $\Omega_{\rm gw}$. 
This quantity serves to capture the ratio of \ac{GW} energy density ${\rm d}\rho_{\rm gw}$ within a specific frequency range [$f$,$f+{\rm d}f$] to the critical energy density $\rho_{\rm c}$, exhibiting a direct relationship with $S_{\rm h}$~\cite{Allen:1996vm,Thrane:2013oya}:
\be
\label{eq:omega_gw}
\Omega_{\rm gw}(f)=\frac{1}{\rho_{\rm c}}\frac{{\rm d}\rho_{\rm gw}}{{\rm d}(\ln{f})}=\frac{2\pi^{2}}{3H_{0}^{2}}f^{3}S_{\rm h}(f),
\ee
where the critical energy density is defined as $\rho_{\rm c}=3H_{0}^{2}c^{2}/(8\pi G)$, with the gravitational constant $G$, and the Hubble constant $H_{0}$. 

\subsection{Overlap reduction function}
The \ac{SGWB} signal $h_{I}(t)$, observed in detector channel $I$, can be expressed as the convolution of the metric perturbations $h(t,\vec{x})$ and the detector response $\mathbb{D}^{ab}(t,\vec{x})$~\cite{Romano:2016dpx}. 
To simplify the measurement process without compromising accuracy, one can focus on a narrow time scale, namely $[t_{0}-T/2,t_{0}+T/2]$, during which the assumption can be made that the channel response remains unchanged. 
By employing the short-term Fourier transform within this time interval, we have: 
\bea
\label{eq:ht_sgwb}
\nn
h_{I}(t,t_{0})&=&\mathbb{D}_{I}[t,\vec{x}(t_{0})]*h[t,\vec{x}(t_{0})]\\
\nn
&=&\sum_{P=+,\times}\int_{-\infty}^{\infty}{\rm d}f
\int_{S^{2}}{\rm d}\hat{\Omega}_{\hat{k}}
F_{I}^{P}(f,\hat{k},t_{0})\widetilde{h}_{P}(f,\hat{k})\\
&&\times e^{{\rm i}2\pi f[t-\hat{k}\cdot\vec{x}(t_{0})/c]},
\eea
where the position vector of the measurement at time $t$ is denoted by $\vec{x}$, the channel response is represented as the double contraction of the channel tensor and the polarization tensor~\cite{Cornish:2001qi}. 

In terms of the frequency domain \ac{SGWB} signal 
\bea
\label{eq:hf_sgwb}
\nn
\widetilde{h}_{I}(f,t_{0})
&=&\sum_{P=+,\times}\int_{S^{2}}\,{\rm{d}}\hat{\Omega}_{\hat{k}}
F_{I}^{P}(f,\hat{k},t_{0})\widetilde{h}_{P}(f,\hat{k}) \\
&&\times e^{-{\rm i}2\pi f\hat{k}\cdot\vec{x}(t_{0})/c},
\eea
the \ac{PSD} of \ac{SGWB} signals is determined by
\be
\label{eq:hIhJ}
\langle\widetilde{h}_{I}(f,t_{0})\widetilde{h}_{J}^{*}(f',t_{0})\rangle
=\frac{1}{2}\delta(f-f')\Gamma_{IJ}(f,t_{0})S_{\rm h}(|f|).
\ee
Here, $I=J$ corresponds to the auto \ac{PSD} of one channel, while $I\neq J$ indicates the cross \ac{PSD} of two channels. 
Combined with~\eq{eq:Ph}, \eq{eq:S2P}, \eq{eq:hf_sgwb}, and~\eq{eq:hIhJ}, the \ac{ORF} 
\be
\label{eq:Gamma_IJ}
\Gamma_{IJ}(f,t_{0})=
\frac{1}{4\pi}\int_{S^{2}}{\rm d}\hat{\Omega}_{\hat{k}}
\mathcal{Y}_{IJ}(f,\hat{k},t_{0}),
\ee
where the antenna pattern involves both the channel response and the separation vector $\Delta \vec{x}=\vec{x}_{I}-\vec{x}_{J}$ between detectors:
\bea
\label{eq:Y_IJ_def}
\nn
\mathcal{Y}_{IJ}(f,\hat{k},t_{0})&=&\frac{1}{2}\sum_{P=+,\times}
F^{P}_{I}(f,\hat{k},t_{0})F^{P*}_{J}(f,\hat{k},t_{0})\\
&\quad&\times
e^{-{\rm i}2\pi f\hat{k}\cdot[\vec{x}_{I}(t_{0})-\vec{x}_{J}(t_{0})]/c},
\eea
Note that, for triangular-shaped detectors, it is convenient to designate one of the vertices as the reference point for the detector position, facilitating the determination of the separation vector between two detectors~\cite{Liang:2022ufy}.

\subsection{Detection sensitivity}
The data $s_{I}$ is the linear addition of the \ac{SGWB} signal $h_{I}$ and the noise $n_{I}$. 
For channels $I$ and $J$ from two independent detectors, we can define their cross-correlation as~\cite{Allen:1997ad}:
\be
\label{eq:S1_IJ}
S'_{IJ}
=\int_{t_{0}-T/2}^{t_{0}+T/2}{\rm d}t
\int_{t_{0}-T/2}^{t_{0}+T/2}{\rm d}t'
s_{I}(t)s_{J}(t')Q_{IJ}(t,t'),
\ee
where $Q_{IJ}$ is a filter function that can be determined in the aim to maximize the \ac{SNR} of the \ac{SGWB}.
Notice though, for reasons that we will explain in detail later, when the noises from channels $I$ and $J$ have non-zero correlation, then the above definition of channel correlation needs to be extended as \cite{Smith:2019wny}:
\bw
\be
\label{eq:S_IJ}
S_{IJ}
=\int_{t_{0}-T/2}^{t_{0}+T/2}{\rm d}t
\int_{t_{0}-T/2}^{t_{0}+T/2}{\rm d}t'
\left[s_{I}(t)s_{J}(t')-\langle n_{I}(t)n_{J}(t')\rangle\right]Q_{IJ}(t,t'),
\ee
\ew
where
\be
\label{eq:nI_nJ}
\langle n_{I}(t)n_{J}(t')\rangle=
\frac{1}{2}\int_{-\infty}^{\infty}{\rm d}f\,
e^{-i2\pi f (t-t')}P_{{\rm n}_{IJ}}(|f|).
\ee
For uncorrelated channels, $\langle n_{I}(t)n_{J}(t')\rangle=0$, so \eq{eq:S_IJ} reduces to \eq{eq:S1_IJ}.
In real data analysis, the noise \ac{PSD} needs to be determined through the observation data, which inevitably contains random fluctuation. Fortunately, by utilizing the smooth and stationary property of the noise \ac{PSD}, we can greatly reduce its random fluctuation by adopting the averaged periodogram method ~\cite{1161901,Nitz:2018rgo,DalCanton:2020vpm}.

Integrating~\eq{eq:S1_IJ} into~\eq{eq:S_IJ} is a straightforward process that unifies the measurements obtained by the cross-correlation and null-channel methods. In the subsequent analysis, we will examine the \ac{SGWB} detection using the unified measurement $S_{IJ}$ for both methods. 
To further streamline subsequent analysis, we designate the detectors as $\mathcal{A}$ and $\mathcal{B}$. 
Assuming that the employed channels in each detector share the same auto-correlation \ac{ORF} $\mathcal{R}_{\mathcal{A}}$ and noise \ac{PSD}, the total \ac{SNR} in the weak-signal limit~\cite{Seto:2020mfd,Liang:2021bde}
\be
\label{eq:snr_lim}
\rho_{\rm tot}=
\left[\frac{2\,T_{\rm tot}}{1+\delta_{\mathcal{A}\mathcal{B}}}
\int_{f_{\rm min}}^{f_{\rm max}}{\rm d}f\,
\frac{\bar{\Gamma}^{2}_{\rm tot}(f)\Omega_{\rm gw}^{2}(f)}
{\hat{D}_{\mathcal{A}\mathcal{B}}(f)}\right]^{1/2},
\ee
where $T_{\rm tot}$ and $[f_{\rm min},f_{\rm max}]$ denote the total correlation time and the detection band, respectively. 
The total time-averaged \ac{ORF}
\be
\bar{\Gamma}_{\rm tot}(f)=
\sqrt{\frac{1}{T_{\rm tot}}\int_{0}^{T_{\rm tot}}{\rm d}t_{0}\,\sum_{IJ}|\Gamma_{IJ}(f,t_{0})|^{2}}.
\ee
The denominator of the integral term
\be
\hat{D}_{\mathcal{A}\mathcal{B}}(f)=\mathcal{R}_{\mathcal{A}}(f)\mathcal{R}_{\mathcal{B}}(f)
\Omega_{{\rm n}_{\mathcal{A}}}(f)\Omega_{{\rm n}_{\mathcal{B}}}(f),
\ee
with the noise energy spectrum density
\be
\label{eq:omega_t}
\Omega_{{\rm n}_{\mathcal{A}}}(f)=
\frac{2\pi^{2}}{3H_{0}^{2}}f^{3}
\frac{P_{{\rm n}_{\mathcal{A}}}(f)}{\mathcal{R}_{\mathcal{A}}(f)}.
\ee

\begin{figure}[t]
	\centering
	\includegraphics[height=5.7cm]{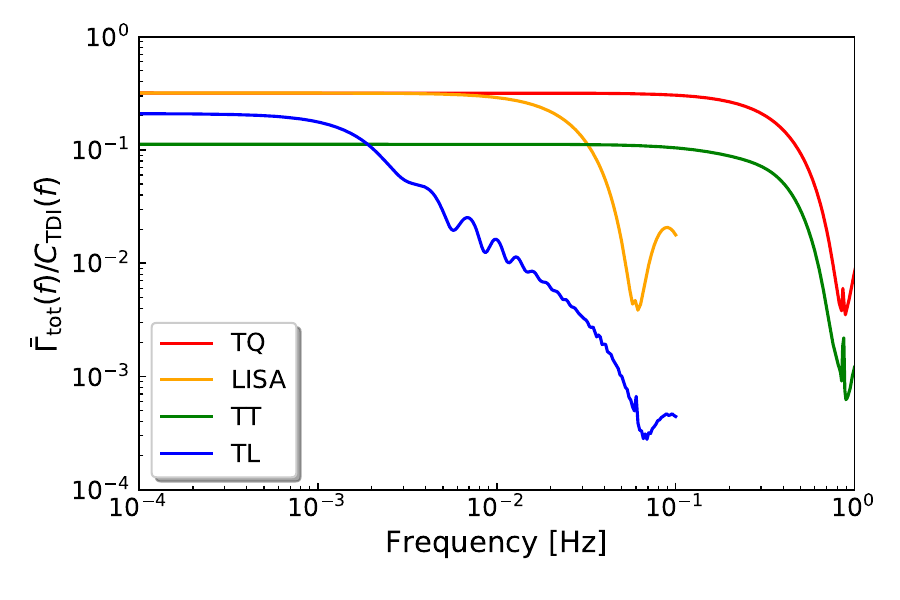}
	\caption{Normalized \ac{ORF} for TianQin (red), LISA (orange), the TianQin I+II network (green), and the TianQin + LISA network (blue).}
	\label{fig:ORF2C}
\end{figure}

Figure~\ref{fig:ORF2C} presents the \ac{ORF} for TianQin, LISA, the TianQin I+II network, and the TianQin + LISA network, employing the A/E channels.~To account for the frequency-dependent factor introduced by the \ac{TDI}, the \ac{ORF} is normalized by $C_{\rm TDI}(f)=4\sin\left[2\pi fL/c\right]\sin\left[2\pi fL'/c\right]$, where $L=L'$ for a single detector~\cite{Cornish:2001bb,Liang:2023fdf}. 
It is observed that the total \ac{ORF} for cross-correlation between two detectors is much lower than that for the auto-correlation of each respective detector channel within the mHz frequency band. 
This effect becomes notably pronounced for the TianQin + LISA network, where the large separation results in a considerable decrease in the drop frequency of \ac{ORF}. 
Furthermore, it is worth mentioning that for the A/E channels of the same detector, the cross-correlation \ac{ORF} is equal to 0.

The next step involves the analysis of the sensitivity curve. 
Considering an \ac{SGWB} characterized by the energy density spectrum that follows a power-law distribution:
\be
\label{eq:Omegaform}
\Omega_{\rm gw}(f)=\Omega_{0}(\epsilon)(f/f_{\rm ref})^{\epsilon}|_{\epsilon=\epsilon_{0}}, 
\ee
the \ac{SNR} threshold $\rho_{\rm thr}$, beyond which the
\ac{GW} detectors can find the \ac{SGWB}, can be translated into the threshold of \ac{SGWB} strength $\Omega_{0}$ based on~\eq{eq:snr_lim}:
\be
\Omega_{0}(\epsilon)=\rho_{\rm thr}\left[\frac{2\,T_{\rm tot}}{1+\delta_{\mathcal{A}\mathcal{B}}}
\int_{f_{\rm min}}^{f_{\rm max}}{\rm d}f
\frac{\bar{\Gamma}^{2}_{\rm tot}(f)(f/f_{\rm ref})^{2\epsilon}}{\hat{D}_{\mathcal{A}\mathcal{B}}(f)}\right]^{-1/2},
\ee
where the reference frequency $f_{\rm ref}$ is arbitrary. 
Subsequently, by considering each frequency with a specific index $\epsilon$, the \ac{PLIS} curve can be generated by
\be
\label{eq:Omega_PLI}
\Omega_{\rm PLIS}(f)={\rm max}_{\epsilon}[\Omega_{0}(\epsilon)(f/f_{\rm ref})^{\epsilon}].
\ee
This sensitivity curve, typically plotted on a logarithmic scale, 
characterizes the collective envelopes of power-law \acp{SGWB} where the \ac{SNR} aligns with the preset threshold. 
Detection expectations for an \ac{SGWB} hinge on its power-law energy spectrum density in relation to the \ac{PLIS} curve. 
If the spectrum surpasses the \ac{PLIS} curve, detection for the \ac{SGWB} is anticipated; however, if it falls below the curve, detection feasibility may be limited by the existing detector configuration.

In addition to following a power-law form, the \ac{SGWB} can also adhere to other forms, such as the first-order \ac{PT}, where the energy spectral density
\be
\Omega_{\rm gw}(f)=\tilde{\Omega}(\boldsymbol{\theta})\mathcal{S}(f,\tilde{f}).
\ee
Here, $\tilde{\Omega}(\boldsymbol{\theta})$ signifies the peak energy spectral density at the peak frequency $\tilde{f}$, with the spectral function $\mathcal{S}(f,\tilde{f})$ being contingent upon the cosmological model. 
To address this disparity, Schmitz et.al introduced the \ac{PIS} curve~\cite{Schmitz:2020syl}:
\be
\label{eq:omega_p}
\Omega_{\rm PIS}(\tilde{f})=\left[\frac{2\,T_{\rm tot}}{1+\delta_{\mathcal{A}\mathcal{B}}}
\int_{f_{\rm min}}^{f_{\rm max}}{\rm d}f
\frac{\bar{\Gamma}^{2}_{IJ}(f)\mathcal{S}^{2}(f,\tilde{f})}
{\hat{D}_{\mathcal{A}\mathcal{B}}(f)}\right]^{-1/2}.
\ee 
Returning to~\eq{eq:snr_lim}, the \ac{SNR} can be determined by
\be
\rho=\frac{\tilde{\Omega}(\boldsymbol{\theta})}{\Omega_{\rm PIS}(\tilde{f})}. 
\ee
By selecting a particular set of parameters $\boldsymbol{\theta}$, the peak frequency $\tilde{f}$ and peak energy spectrum density $\tilde{\Omega}(\boldsymbol{\theta})$ are specified. 
If the peak amplitude exceeds $\rho_{\rm thr}$ times the \ac{PIS} curve, the detection of cosmological \ac{PT} can be confirmed.

\section{Beyond weak-signal limit}\label{sec:beyond}
The weak-signal limit is applicable in most scenarios, especially for the derivation of the sensitivity curves. 
However, in certain cases, the intensity of a \ac{SGWB} could become comparable to or even higher than the detector noise. 
For instance, the foreground originating from \ac{GDWD} can exceed the noise level of space-borne detector noise~\cite{Boileau:2021sni}, and the foreground from \acp{BBH} can dominate over the noise of next-generation ground-based detectors~\cite{Regimbau:2016ike}. 
In such cases, adopting the \ac{SNR} beyond the weak-signal limit becomes necessary.

Assuming that all other conditions specified in~\eq{eq:S_IJ} hold true, except for the weak-signal limit. 
Then for all channel pairs $\{IJ\}$ (or equivalently, $\{MN\}$), the expectation and variance of the detection measurement
\bw
\bea
\label{eq:mu_IJ}
\nn
\mu&=&\langle \sum_{IJ}S_{IJ} \rangle\\
\nn
&=&\sum_{IJ}\int_{0}^{T_{\rm tot}}{\rm d}t_{0}
\int_{-\infty}^{\infty}{\rm d}f
\int_{-\infty}^{\infty}{\rm d}f'
\,\langle\widetilde{h}_{I}(f,t_{0})\widetilde{h}^{*}_{J}(f',t_{0})\rangle \widetilde{Q}_{IJ}(f',t_{0})
e^{-{\rm i}2\pi (f-f')t_{0}} \\
\nn
&=&
\sum_{IJ}
\int_{0}^{T_{\rm tot}}{\rm d}t_{0}
\int_{0}^{\infty}{\rm d}f\int_{0}^{\infty}{\rm d}f'\,
\delta(f-f')\Gamma_{IJ}(f,t_{0})S_{\rm h}(f)\widetilde{Q}_{IJ}(f',t_{0})e^{-{\rm i}2\pi(f-f')t_{0}}\\
&=&
\sum_{IJ}
\frac{3H_{0}^{2}}{2\pi^{2}}\int_{0}^{T_{\rm tot}}{\rm d}t_{0}
\int_{0}^{\infty}{\rm d}f\,
f^{-3}\Gamma_{IJ}(f,t_{0})\Omega_{\rm gw}(f)
\widetilde{Q}_{IJ}(f,t_{0}),
\eea
\bea
\label{eq:sigma_IJ}
\nn
\sigma^{2}&=&\langle \mu^{2}\rangle-\langle \mu\rangle^{2}\\
\nn
&=&\sum_{IJ,MN}
\int_{0}^{T_{\rm tot}}{\rm d}t_{0}
\int_{0}^{T_{\rm tot}}{\rm d}\eta_{0}
\int_{-\infty}^{\infty}{\rm d}f\int_{-\infty}^{\infty}{\rm d}f'
\int_{-\infty}^{\infty}{\rm d}\omega\int_{-\infty}^{\infty}{\rm d}\omega'
\\
\nn
&&\times
\bigg[\left\langle\big(\widetilde{s}_{I}(f,t_{0})\widetilde{s}_{J}^{*}(f',t_{0})
-\langle\widetilde{n}_{I}(f,t_{0})\widetilde{n}_{J}^{*}(f',t_{0})\rangle\big)
\big(\widetilde{s}_{M}(\omega,\eta_{0})\widetilde{s}_{N}^{*}(\omega',\eta_{0})
-\langle\widetilde{n}_{M}(\omega,\eta_{0})\widetilde{n}_{N}^{*}(\omega',\eta_{0})\rangle\big)
\right\rangle\\
\nn
&&-\langle \widetilde{h}_{I}(f,t_{0})\widetilde{h}_{J}^{*}(f',t_{0}) \rangle
\langle \widetilde{h}_{M}(\omega,\eta_{0})\widetilde{h}_{N}^{*}(\omega',\eta_{0}) \rangle\bigg]
\widetilde{Q}_{IJ}(f',t_{0})\widetilde{Q}^{*}_{MN}(\omega',\eta_{0})
e^{-{\rm i}2\pi(f-f')t_{0}}
e^{-{\rm i}2\pi(\omega-\omega')\eta_{0}}\\
\nn
&=&
2\sum_{IJ,MN}
\int_{0}^{T_{\rm tot}}{\rm d}t_{0}
\int_{0}^{\infty}{\rm d}f\,
\bigg[
\langle \widetilde{s}_{I}(f,t_{0})\widetilde{s}_{M}(f,t_{0}) \rangle
\langle \widetilde{s}_{J}(f,t_{0})\widetilde{s}_{N}(f,t_{0}) \rangle
+
\langle \widetilde{s}_{I}(f,t_{0})\widetilde{s}_{N}(f,t_{0}) \rangle
\langle \widetilde{s}_{J}(f,t_{0})\widetilde{s}_{M}(f,t_{0}) \rangle
\bigg]\\
\nn
&&\times
\widetilde{Q}_{IJ}(f,t_{0})\widetilde{Q}^{*}_{MN}(f,t_{0})
\\
&\approx&
\frac{1+\delta_{\mathcal{A}\mathcal{B}}}{2}
\left[\frac{3H_{0}^{2}}{2\pi^{2}}\right]^{2}
\sum_{IJ}
\int_{0}^{T_{\rm tot}}{\rm d}t_{0}
\int_{0}^{\infty}{\rm d}f\,
f^{-6}D_{\mathcal{A}\mathcal{B}}(f)|\widetilde{Q}_{IJ}(f,t_{0})|^{2},
\eea
where the filter function $\widetilde{Q}_{IJ}$ is arbitrary. 
Furthermore, for the last line of \eq{eq:sigma_IJ} to be valid, it is necessary that the \ac{ORF} for cross-correlation is significantly lower than that for auto-correlation, leading to
\be
\label{eq:D_snr}
D_{\mathcal{A}\mathcal{B}}(f)=
\mathcal{R}_{\mathcal{A}}(f)\mathcal{R}_{\mathcal{B}}(f)
\left[\Omega_{{\rm n}_{\mathcal{A}}}(f)+\Omega_{\rm gw}(f)\right]
\left[\Omega_{{\rm n}_{\mathcal{B}}}(f)+\Omega_{\rm gw}(f)\right].
\ee

Combined with \eq{eq:mu_IJ} and \eq{eq:sigma_IJ}, the square of \ac{SNR} can be obtained by
\be
\label{eq:snr_sgwb}
\rho^{2}=\frac{\mu^{2}}{\sigma^{2}}=
\frac{2}{1+\delta_{\mathcal{A}\mathcal{B}}}
\frac{\big[\sum_{IJ}\int_{0}^{T_{\rm tot}}{\rm d}t_{0}
\int_{0}^{\infty}{\rm d}f\,
f^{-3}\Gamma_{IJ}(f,t_{0})\Omega_{\rm gw}(f)\widetilde{Q}_{IJ}(f,t_{0})\big]^{2}}
{\sum_{IJ}\int_{0}^{T_{\rm tot}}{\rm d}t_{0}
\int_{0}^{\infty}{\rm d}f\,
f^{-6}D_{\mathcal{A}\mathcal{B}}(f)|\widetilde{Q}_{IJ}(f,t_{0})|^{2}}.
\ee
\ew
From~\eq{eq:snr_sgwb}, it is evident that the extension from~\eq{eq:S1_IJ} to~\eq{eq:S_IJ} guarantees an \ac{SNR} of 0 in the absence of an \ac{SGWB} signal.

\begin{figure}[ht]
	\centering
	\includegraphics[height=7cm]{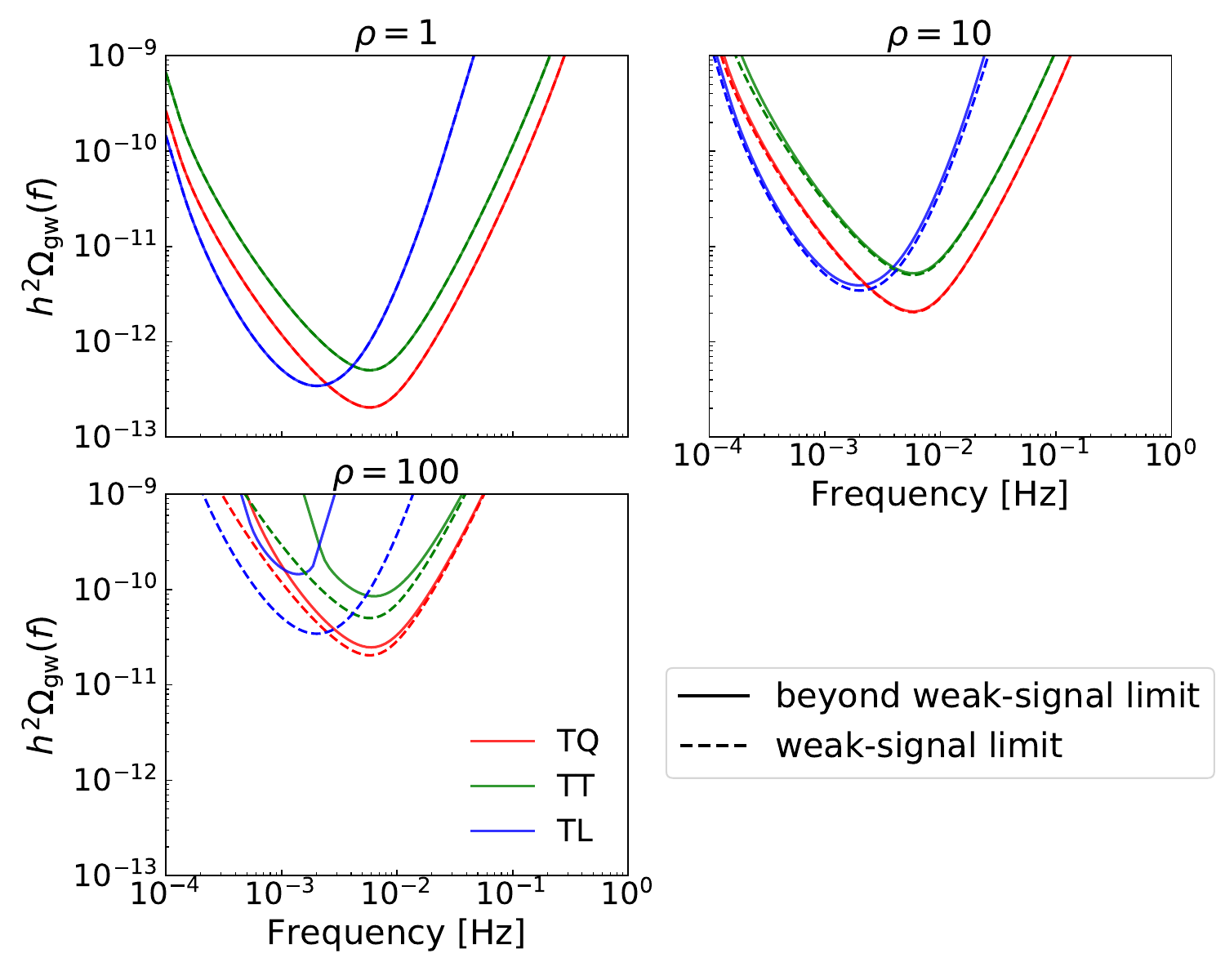}
	\caption{\ac{PLIS} curve with and without weak-signal limit.}
	\label{fig:Omega_PI}
\end{figure}

\begin{figure}[ht]
	\centering
	\includegraphics[height=7cm]{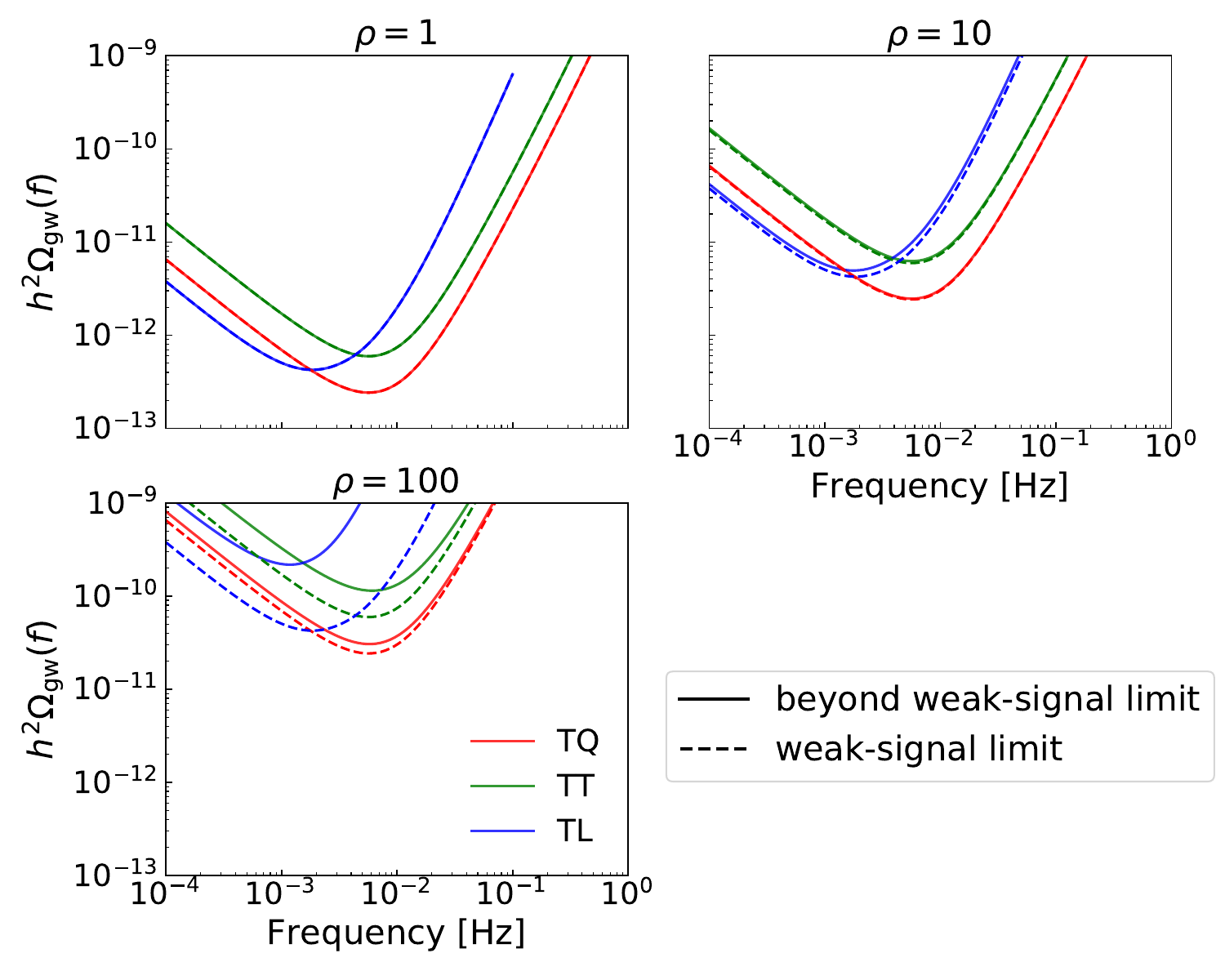}
	\caption{\ac{PIS} curve with and without weak-signal limit.}
	\label{fig:Omega_P}
\end{figure}

According to the definition of a positive-definite inner product, which is applicable to any pair of complex functions $A$ and $B$~\cite{Allen:1997ad}:
\be
(A,B):=\sum_{IJ}\int_{0}^{T_{\rm tot}}{\rm d}t_{0}
\int_{0}^{\infty}{\rm d}f\,
A_{IJ}(f,t_{0})B_{IJ}^{*}(f,t_{0})D_{\mathcal{A}\mathcal{B}}(f),
\ee
\eq{eq:snr_sgwb} transitions to 
\be
\rho_{\rm opt}^{2}=
\frac{2}{1+\delta_{\mathcal{A}\mathcal{B}}}
\frac{\big(\widetilde{Q}_{IJ}(f,t_{0}),\frac{f^{3}\Gamma^{*}_{IJ}(f,t_{0})\Omega_{\rm gw}(f)}
{D_{\mathcal{A}\mathcal{B}}(f)}\big)^{2}}{\big(\widetilde{Q}_{IJ}(f,t_{0}),\widetilde{Q}_{IJ}(f,t_{0})\big)}.
\ee
Maximizing the \ac{SNR} yields the solution
\be
\widetilde{Q}_{IJ}(f,t_{0})
=\lambda\frac{f^{3}\Gamma^{*}_{IJ}(f,t_{0})\Omega_{\rm gw}(f)}{D_{\mathcal{A}\mathcal{B}}(f)},
\ee
with a real constant $\lambda$. Following the preceding derivation, the total \ac{SNR} should be modified by
\be
\label{eq:snr_tot}
\rho_{\rm tot}=
\sqrt{\frac{2\,T_{\rm tot}}{1+\delta_{\mathcal{A}\mathcal{B}}}
\int_{f_{\rm min}}^{f_{\rm max}}{\rm d}f\,
\frac{\bar{\Gamma}^{2}_{\rm tot}(f)\Omega_{\rm gw}^{2}(f)}
{D_{\mathcal{A}\mathcal{B}}(f)}}.
\ee
In contrast to~\eq{eq:snr_lim},~\eq{eq:snr_tot} suggests that the \ac{SGWB} detection can suffer negative impacts from the \ac{SGWB} itself, especially in cases of loud \ac{SGWB}. 
To visualize this impact, plotting sensitivity curves is beneficial. 
Although a direct derivation of $\Omega_{0}$ and $\Omega_{\rm PIS}$ directly from \eq{eq:Omegaform} and \eq{eq:omega_p} based on~\eq{eq:snr_tot} to generate the sensitivity curves proves challenging, these values can still be determined by inverting the \ac{SNR} formula\footnote{In this work, we utilize the Newton-Raphson method.}. 
\fig{fig:Omega_PI} and~\fig{fig:Omega_P} illustrate the \ac{PLIS} and \ac{PIS} curves correspondingly, with solid lines in red, green, and blue denoting TianQin, the TianQin I+II network, and the TianQin + LISA network~\footnote{Note that for the TianQin detector, the auto-correlations within the same channel are considered. In the case of detector networks, the cross-correlations between channels from different detectors are taken into consideration.}, surpassing the weak-signal limit. 
Dashed lines including the weak-signal limit are provided for comparison. 
As an illustrative example, the spectral function for the \ac{PIS} curves is determined by
\be
S(f,\tilde{f}_{i})
=\frac{3.9(f/\tilde{f}_{i})^{2.9}}{1+2.9(f/\tilde{f}_{i})^{3.8}}.
\ee
Furthermore, the operation time is set at 1 year, resulting in a correlation time $T_{\rm tot}$ equal to half a year, four months, and half a year for TianQin, the TianQin I+II network, and the TianQin + LISA network, respectively.

While sensitivity curves exhibit minimal gaps with and without the weak-signal limit when the \ac{SNR} remains below 10, 
a notable gap emerges as the \ac{SNR} reaches 100, especially pronounced for the TianQin + LISA network, with gaps reaching up to an order of magnitude. 
This suggests that the energy spectral density predicted by the weak-signal limit needs to be raised by approximately an order of magnitude for detectability.

\section{Case study}\label{sec:Case}
In this section, we will briefly discuss the potential sources that could generate loud \ac{SGWB} in the mHz range, including \acp{DWD} and first-order \acp{PT}. 

\acp{GDWD} are anticipated to produce foreground comparable to the noise of space-borne detectors. In terms of our previous work, the \ac{PSD} of Galactic foreground can be fit for through the following polynomial~\cite{Huang:2020rjf,Liang:2021bde}:
\be
S_{\rm h}(f)=\frac{20}{3}\left[10^{\sum_i a_{i}x^{i}}\right]^{2},
\ee
where $x=\log(f/10^{-3}\,\,{\rm Hz})$, $a_{i}$ represents the polynomial coefficients that are updated based on different operation times. The specific values of these coefficients can be found in~\tab{tab:DWD}. 
In terms of~\eq{eq:omega_gw}, the energy spectrum density of Galactic foreground can be further derived. 

\begin{table}[t]
	\begin{center}
		\caption{Coefficients for the polynomial fit for the Galactic foreground. Each successive row corresponds to an increasing operation time $T$.}
		\centering
		\setlength{\tabcolsep}{1.4mm}
		\renewcommand\arraystretch{1.5}
		\label{tab:DWD}
		\begin{tabular}{c c c c c c c c}
			\hline
			\hline
			$T$   &$a_{0}$ & $a_{1}$&$a_{2}$ &$a_{3}$& $a_{4}$&$a_{5}$ &$a_{6}$\\
			\hline
			0.5 yr  & -18.7  & -1.23   & -0.801  &  0.832  &  -1.96   &  3.09  & -2.38\\
			1 yr    & -18.7  & -1.34   & -0.513  &  0.0152 &  -1.53   &  4.79  & -5.01\\
			2 yr    & -18.7  & -1.39   & -0.610  &  0.577  & 0.00242 &  0.578 & -4.39 \\
			4 yr    & -18.7  & -1.30   & -0.872  &  0.266  & -5.12  & 15.6  & -15.5\\
			5 yr    & -18.7  & -1.32   & -0.322  & -1.68   & -4.49  &  21.6 & -22.6\\
			\hline
			\hline
		\end{tabular}
	\end{center}
\end{table}

\begin{table*}
	\begin{center}
		\caption{Ingredients contributing to the \ac{SGWB} spectra arising from first-order \acp{PT}. The parameter
		$\alpha_{\infty}$ determines the threshold for the weakest transition at which a portion of the vacuum energy converts into kinetic energy, driving the expansion of bubbles~\cite{Ellis:2019oqb}. The sound speed is given by $c_{\rm s}=1/\sqrt{3}$, the redshifted, present-day value of the Hubble frequency $h_{*}=1.65\times10^{-5}\left(\frac{T_{*}}{100 {\rm~GeV}}\right)\left(\frac{g_{*}}{100 {\rm~GeV}}\right)^{1/6}${\rm~Hz}.}
		\label{tab:para_pt}
		\setlength{\tabcolsep}{3.6mm}
		\renewcommand\arraystretch{2}
		{
			\begin{tabular}{c c c c c c c}
				\hline
				\hline
				$i$         & $p$  & $q$  & $\tilde{\Delta}_{i*} (v_w)$  & $\frac{\tilde{f}_{i*}}{\beta}$ 
				&  $\mathcal{S}_{i}(f,\tilde{f}_{i})$ &  $K_{i}(\alpha)$       \cr 
				\hline
				col       & 2  & 2 & $\frac{0.44v_{w}^{3}}{1+8.28v_{w}^{3}}$  & $\frac{0.31}{1-0.051v_{w}+0.88v_{w}^{2}}$ 
				&  $\left(\frac{f}{\tilde{f}_{i}}\right)^{2.8}\frac{3.8}{1+2.8(f/\tilde{f}_{i})^{3.8}}$ &  ${\rm max}\big[1-\frac{\alpha_{\infty}}{\alpha},0\big]$     \cr
				\multirow{1}{*}{sw} & 1  & 2 & $0.157v_{w}H_{*}\tau_{\rm sw}$  & $1.16(1-c_{\rm s}/v_{w})$ &   $\left(\frac{f}{\tilde{f}_{i}}\right)^{3}\left[\frac{7}{4+3(f/\tilde{f}_{i})^{2}}\right]^{7/2}$    &  
				\multirow{2}{*}{$\left\{\begin{gathered}
						\kappa(\alpha_{N})|_{\alpha_{N}=\alpha}, \alpha\le\alpha_{\infty}\\
						\frac{\alpha_{\infty}}{\alpha}\kappa(\alpha_{N})|_{\alpha_{N}=\alpha_{\infty}},\alpha>\alpha_{\infty}
					\end{gathered}
					\right.$}  \cr
				\multirow{1}{*}{turb}       & 1  & $\frac{3}{2}$ & $20v_{w}(1-H_{*}\tau_{\rm sw})$  & $1.33(1-c_{\rm s}/v_{w})$
				& $\left(\frac{f}{\tilde{f}_{i}}\right)^{3}\frac{[1+(f/\tilde{f}_{i})]^{3/11}}{1+8\pi f/h_{*}}$  &    \cr
				\hline
				\hline
		\end{tabular}}
	\end{center}
\end{table*}

Furthermore, it has been highlighted that the \ac{SGWB} originating from \acp{EDWD} may exhibit significant intensity levels, leading to high \ac{SNR}~\cite{Liang:2021bde}. 
Without loss of generality, the corresponding energy spectrum density of \ac{SGWB} can be described as the cumulative contribution from each binary system~\cite{Phinney:2001di,LIGOScientific:2016fpe,Romano:2016dpx,LIGOScientific:2017zlf}: 
\be
\label{eq:omegastro}
\Omega_{\rm gw}(f)=\frac{f}{\rho_{\rm c}}
\int\,{\rm d}\boldsymbol{\theta}\int_{0}^{z_{\rm max}}\,{\rm d}z
\frac{R_{\rm m}(z)\frac{{\rm d}E_{\rm gw}}{{\rm d}f_{\rm s}}(f_{\rm s},\boldsymbol{\theta})}{(1+z)H(z)},
\ee
where the population parameter $\boldsymbol{\theta}$ primarily encompasses the component masses of the system. 
Given the rarity of astrophysical compact stars at high redshifts due to the lack of star formation, the Hubble parameter $H(z)$ can be accurately approximated as $H_{0}\sqrt{\Omega_{\rm m}(1+z)^{3}+\Omega_{\Lambda}}$ ~\cite{Planck:2018vyg}. 
In this paper, we focus on the DA and DB white dwarfs, for which the mass distribution can be modeled by a Gaussian distribution with mean $\mu=0.663\,\,M_{\odot}$ and standard deviation $\sigma=0.177\,\,M_{\odot}$~\cite{Rosado:2011kv}. 
The merger rate is directly linked to the redshift and can be expressed as follows~\cite{KAGRA:2021duu}:
\be
R_{\rm m}(z)=R_{0}(1+z)^{\kappa},
\ee
with the local merger rate $R_{0}$ and index $\kappa$. 
As suggested in Ref.~\cite{Rosado:2011kv}, the $R_{0}$ is assumed to be within the range of 20 to $500\,\,{\rm Mpc^{-3}\,Myr^{-1}}$. 
The energy spectrum emitted in the source frame, denoted by ${\rm d}E_{\rm gw}/{\rm d}f_{\rm s}$, is characterized by the frequency in the source frame $f_{\rm s}=f(1+z)$~\cite{Zhu:2012xw}. 
Given that the radius of a white dwarf significantly exceeds its Schwarzschild radius, it is reasonable to assume that below the cut-off frequency where both stars come into contact, the binary systems consistently remain in the inspiral phase.

\begin{figure}
	\centering
	\includegraphics[height=5.8cm]{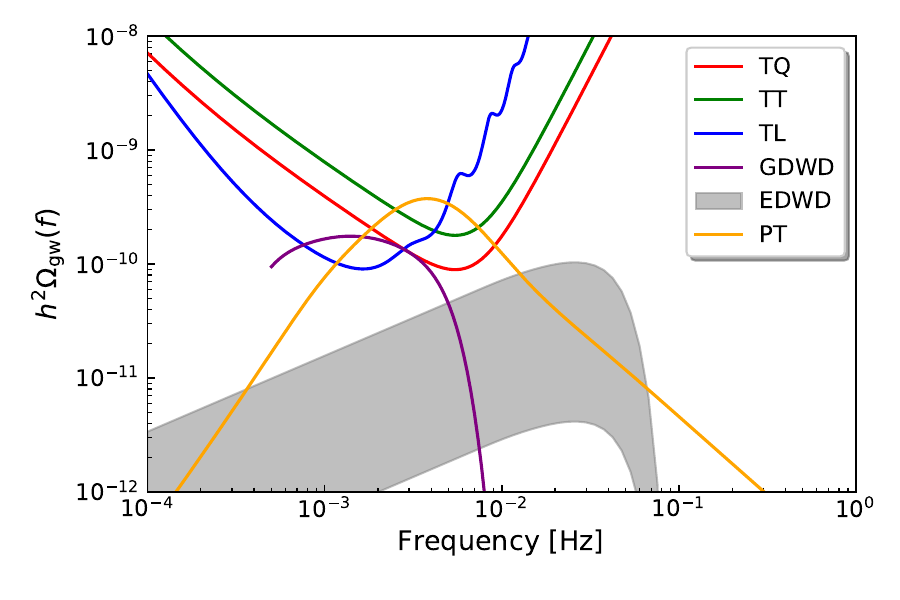}
	\caption{Noise energy spectrum density of TianQin, the TianQin I+II network, the TianQin + LISA network, along with \ac{SGWB} energy spectrum density of \acp{DWD} and first-order \acp{PT}.}
	\label{fig:Omega_all}
\end{figure}

Regarding the first-order \acp{PT}\footnote{Unless otherwise specified, we adopt the natural unit for discussions involving first-order \acp{PT}.}, we adopt the model mentioned in our previous study~\cite{Liang:2021bde}, where the energy spectrum density is formulated as
\be
\label{eq:Omega_PT}
\Omega_{{\rm gw}}(f)=
\sum_{i} \tilde{\Omega}_{i}(\boldsymbol{\theta})\mathcal{S}_{i}(f,\tilde{f}_{i}).
\ee
Here, the sum accounts for the collision occurring during the \ac{PT}, as well as the subsequent acoustic and turbulent phases. 
The redshifted peak frequency today for each source is given by~\cite{Jinno:2016vai}
\be
\tilde{f}_{i}=\frac{\tilde{f}_{i*}}{\beta}
\left(\frac{\beta}{H_{*}}\right)h_{*},
\ee
where the peak frequency before redshifting, denoted as $\tilde{f}_{i*}$, depends on the sound speed $c_{\rm s}$ and the bubble velocity $v_w$. 
The parameter $\beta/H_{*}$ represents \ac{PT} duration normalized to the Hubble parameter, while $h_{*}$ indicates the Hubble frequency. 
Additional details are provided in~\tab{tab:para_pt}. 

When \acp{GW} are generated with the Hubble rate $H_{*}$, temperature $T_{*}$, and the total number of relativistic degrees of freedom $g_{*}$, the individual contributions factoring in the redshift can be succinctly expressed as
\bea
\nn
\tilde{\Omega}_{i}(\boldsymbol{\theta})=\,\,&&
1.67\times10^{-5}\left(\frac{100}{g_{*}(T_{*})}\right)^{1/3}
\left(\frac{H_{*}}{\beta}\right)^{p}\\
&&\times
\left(K_{i}(\alpha ) \frac{\alpha}{1+\alpha}\right)^{q}\tilde{\Delta}_{i*}(v_w),
\label{eq:omegapt}
\eea
where the indices $p,q$, the spectral function $\mathcal{S}_{i}(f,\tilde{f}_{i})$, the efficiency factor $K_{i}$, and the peak amplitude $\tilde{\Delta}_{i*}$ are detailed in \tab{tab:para_pt}. 
It is worth noting that the peak amplitude involves the timescale for acoustic production~\cite{Hindmarsh:2017gnf,Ellis:2019oqb,Wang:2020jrd}
\bea
\tau_{\rm sw}&=&\min \left[{H^{-1}_{*}},R_{*}/U_{\rm f} \right],
\eea
with the mean bubble separation $R_{*}$ and the average square root of the fluid velocity $U_{\rm f}$~\cite{Espinosa:2010hh}. 

\begin{table}[t]
	\begin{center}
		\caption{Detection \ac{SNR} for Galactic foreground.}
		\setlength{\tabcolsep}{6.7mm}
		\renewcommand\arraystretch{2}
		\label{tab:snr_dwd}
		{
			\begin{tabular}{c c c c}
				\hline\hline
				 & TQ & TT & TL \cr 
				\hline
				With limit & 300 & 120 & 430 \cr 
				Without limit & 170 & 69 & 96 \cr 
				\hline
				\hline
			\end{tabular}}
	\end{center}
\end{table}

Figure.~\ref{fig:Omega_all} illustrates the \acp{SGWB} from various sources: the purple line represents the Galactic foreground with a one-year operation time, the gray region corresponds to the extragalactic background associated with the local merger rate ranging from 20 to $500\,\,{\rm Mpc^{-3}\,Myr^{-1}}$, and the orange line signifies the cosmological background determined by the parameters $\{\alpha,\alpha_{\infty},v_{w},\beta/H_{*},T_{*}\}=\{2,1,0.95,50,10^{3}\,{\rm{GeV}}\}$. 
To facilitate comparison between the \ac{SGWB} intensity and noise levels, the noise energy spectrum density of TianQin, the TianQin I+II network, and the TianQin + LISA network are shown in red, green, and blue lines, respectively. 
While the extragalactic background can near the detector noise level, both the Galactic foreground and cosmological background can exceed the noise. 
Taking the Galactic foreground as an example,~\tab{tab:snr_dwd} presents the detection \ac{SNR} within different detector configurations. 
It also provides the \ac{SNR} under weak-signal conditions for the sake of comparison. 
In the weak-signal limit, the TianQin + LISA network achieves the highest \ac{SNR} for detecting the Galactic foreground. 
As moving beyond the weak-signal limit, the \ac{SNR} drops to approximately 20\% for the TianQin + LISA network, whereas it remains around 60\% for TianQin and the TianQin I+II network. Thus, TianQin demonstrates superior performance in terms of \ac{SNR}.

The other two backgrounds exhibit considerable uncertainties in their parameters compared to the Galactic foreground. We will proceed to utilize \ac{SNR} contour plots to determine the detectable parameter regions for them. 

\begin{figure}[t]
	\centering
	\includegraphics[height=5.8cm]{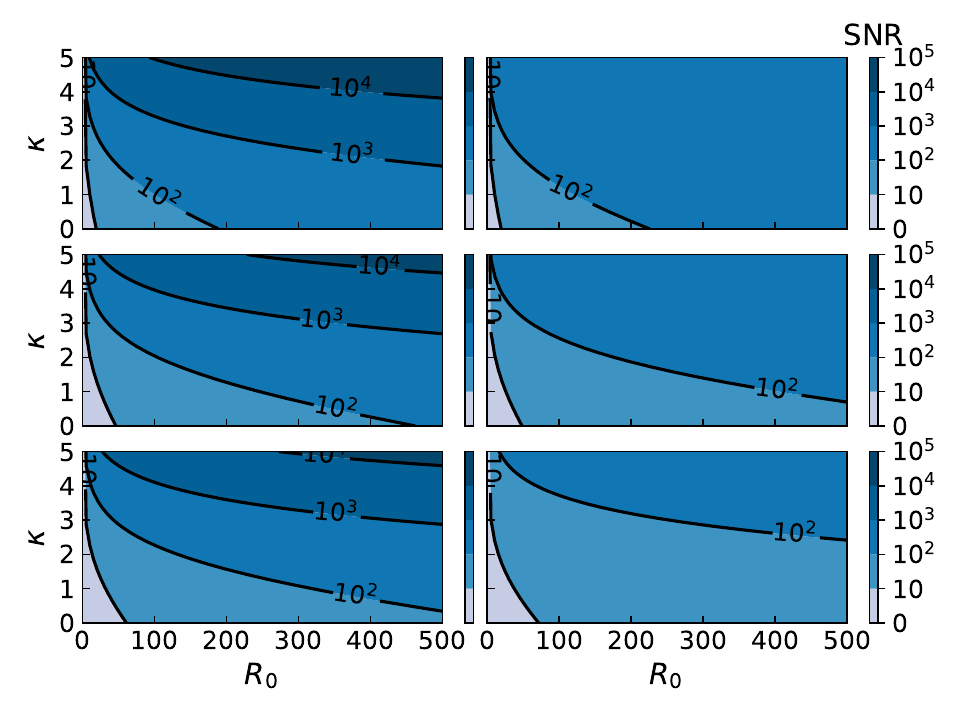}
	\caption{\ac{SNR} contour plots for the \ac{SGWB} resulting from extragalactic \ac{DWD}. The left-hand and right hand panels refer to the scenarios with and without weak-signal limit, considering a one-year operation time.}
	\label{fig:SNR_EDWD}
\end{figure}

In~\fig{fig:SNR_EDWD}, we present \ac{SNR} contour plots for the extragalactic background, showcasing scenarios with and without the weak-signal limit in the left and right panels for comparison. 
Not surprisingly, an \ac{SNR} threshold of 100 leads to significant variations in the detectable parameter regions between the two scenarios, especially for the TianQin + LISA network. 
Failure to achieve an \ac{SNR} above 100 within one year of operation would shift the lower limit for $\kappa$ from around 0 to 2.5, indicating a transition from a scenario with the weak-signal limit to one without it. 
This shift suggests that the adverse effects of loud \ac{SGWB} self-generation on detection reduce the region of detectable parameters to a lesser extent than initially anticipated.  
On the other hand, assuming a constant merger rate with redshift ($\kappa=0$), achieving an \ac{SNR} exceeding 100 after 1 year of operation, under the weak-signal limit, would estimate the merger rate for \acp{EDWD} to be around $500\,\,{\rm Mpc^{-3}\,Myr^{-1}}$. 
However, this value falls considerably short of meeting the actual detection requirements.

\begin{figure}[t]
	\centering
	\includegraphics[height=5.8cm]{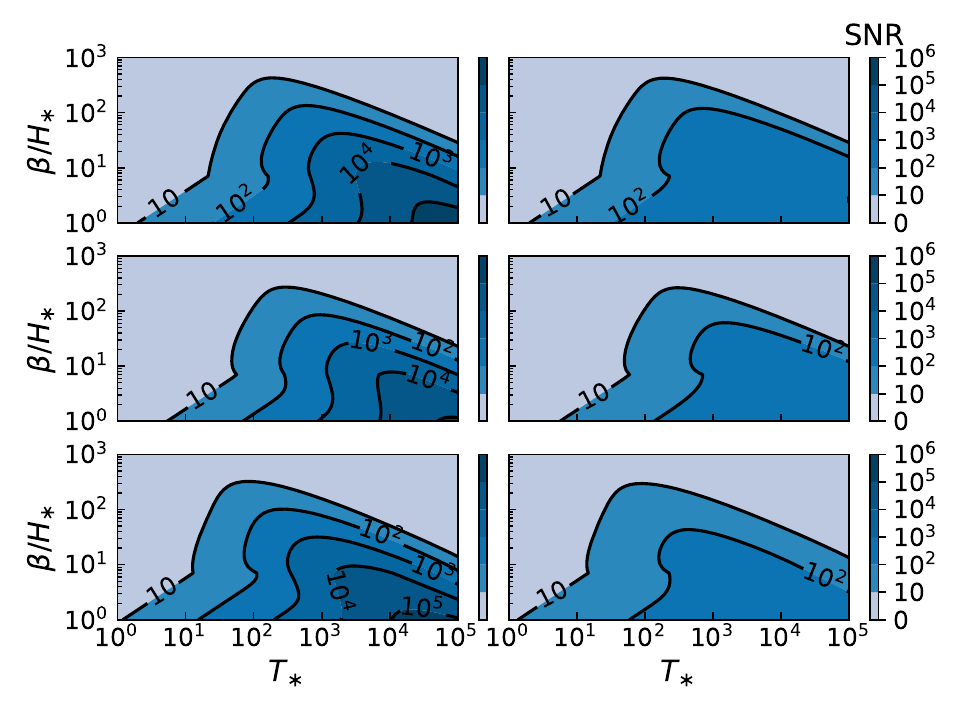}
	\caption{\ac{SNR} contour plots for the \ac{SGWB} resulting from first-order \acp{PT}. The left-hand and right hand panels refer to the scenarios with and without weak-signal limit, considering a one-year operation time.}
	\label{fig:SNR_PT_beta_T}
\end{figure}

A similar diminishing effect of parametric limitations can be observed in~\fig{fig:SNR_PT_beta_T}, where we follow Refs.~\cite{Caprini:2015zlo,Caprini:2019egz} and set the parameters $\{\alpha,\alpha_{\infty},v_{w}\}=\{2,1,0.95\}$ for the first-order \acp{PT}. By setting $\alpha$ greater than $\alpha_{\infty}$, the \ac{SGWB} is anticipated to involve the collision, acoustic, and turbulent phases; the bubble velocity $v_{w}$ exceeding the sound speed $c_{\rm s}$ corresponds to detonation bubble collisions~\cite{Kamionkowski:1993fg}. Note that, changing other \ac{PT} parameters and the parameter degeneracy may also influence the results depicted in the~\fig{fig:SNR_PT_beta_T}. 
If the \ac{SNR} surpasses 100 after 1 year of operation, the temperature $T_{*}$ is expected to be significantly greater than $10\,\,{\rm{GeV}}$, while the \ac{PT} duration $\beta/H_{*}$ should be less than $10^{2}$. 
The detectable region for these two parameters can expand notably by utilizing the weak-signal limit, leading to potential misjudgments in detection. 
It is important to highlight that, due to the wide range of variability in fixed parameters, effectively constraining the parameters of \acp{PT} remains a challenging task. 

\section{Conclusions and discussions}\label{sec:conclusion}
In this paper, we have derived the \ac{SNR} tailored for detecting loud \ac{SGWB}. 
By utilizing the improved \ac{SNR} estimation, we have plotted \ac{PLIS} and \ac{PIS} curves for TianQin, the TianQin I+II network, and TianQin + LISA network to assess the detectable \ac{SGWB} energy spectrum density over an operational period of 1 year. 
Our findings indicated that for \ac{SNR} values below 10, the impact of considering or neglecting the weak-signal limit on the results is minimal. 
However, when the \ac{SNR} exceeds 100, applying the weak-signal limit can lead to a considerable overestimation of the detectable energy spectrum density. 
This overestimation becomes markedly pronounced for the TianQin + LISA network, where it may be inflated by approximately an order of magnitude.

Additionally, we have employed \ac{SNR} contour plots to determine the detectable parameter regions for loud \acp{SGWB} generated by \acp{EDWD} and first-order \acp{PT}. 
With a chosen \ac{SNR} threshold of 100, the prospect for detecting the \ac{SGWB} from \acp{EDWD}, specifically DA and DB types, appears promising. 
This detection is grounded in a consistent merger rate exceeding $200\,\,{\rm Mpc^{-3}\,Myr^{-1}}$ surpassing varying redshifts, with the operation period extending over 1 year. 
On the other hand, while it is feasible to narrow down the parameters for first-order \acp{PT} by holding certain variables fixed, uncertainties in the underlying models permit only a broad constriction of the detectable parameter region.

We emphasize that our analysis can extend to the statistical inference of \ac{SGWB} detection, which encompasses classical (frequentist) inference and Bayesian inference. 
The pivotal aspect lies in constructing the likelihood function. 
By estimating the expectation and variance in terms of~\eq{eq:mu_IJ} and \eq{eq:sigma_IJ}, one can ensure that the likelihood function aligns more closely with the actual detection process, thereby minimizing bias in parameter estimation. 
This approach is not limited to space-borne detection~\cite{Caprini:2019pxz,Pieroni:2020rob,Flauger:2020qyi}, but is also relevant for next-generation ground-based detection~\cite{Zhong:2022ylh,Bellie:2023jlq}.

In this work, we assume the ideal case where the only \ac{GW} signal is \ac{SGWB}.
In reality, different signals will overlap and make the analysis more complicated.
Fortunately, even if the \ac{SGWB} is strong, it will not complicate the analysis of other sources.
Taking the foreground from Galactic compact binaries as an example, it is expected that this foreground is comparable to or even stronger than instrument noise.
However, when analyzing signals like the merger of massive black holes, the foreground can just be regarded as part of the \ac{PSD}.
That being said, the strong \ac{SGWB} will inevitable raise the noise level, causing a decrease in the sensitivity.

\begin{acknowledgments}
This work has been supported by the Guangdong Major Project of Basic and Applied Basic Research (Grant No. 2019B030302001), the National Key Research and Development Program of China (No. 2020YFC2201400), the Natural Science Foundation of China (Grants No. 12173104), the Natural Science Foundation of Guangdong Province of China (Grant No. 2022A1515011862), and the Guangdong Basic and Applied Basic Research Foundation(Grant No. 2023A1515030116). 
Z.C.L. is supported by the China Postdoctoral Science Foundation (Grant No. 2023M744094), and the Guangdong Basic and Applied Basic Research Foundation(Grant No. 2023A1515111184). 
We also thank Fapeng Huang, Yun Jiang, and Jianwei Mei for helpful discussions. 
\end{acknowledgments}


\normalem
\bibliographystyle{apsrev4-1}
\bibliography{ref}

\end{document}